\documentclass[aps,prl,twocolumn,showpacs,floatfix,superscriptaddress]{revtex4}
\usepackage{graphicx}
\usepackage{psfrag}

\begin{document}

\title{Chaos and statistical relaxation in quantum systems of interacting particles}

\author{L.~F. Santos}
\email{lsantos2@yu.edu}
\affiliation{Department of Physics, Yeshiva University, 245 Lexington Ave, New York, NY 10016, USA}
\author{F. Borgonovi}
\email{fausto.borgonovi@unicatt.it}
\affiliation{Dipartimento di Matematica e Fisica, Universit\'a Cattolica, via Musei 41, 25121 Brescia, and INFN,
Sezione di Pavia, Italy}
\author{F.~M. Izrailev}
\email{felix.izrailev@gmail.com}
\affiliation{Instituto de F{\'i}sica, Universidad Aut\'onoma de Puebla,
Apt. Postal J-48, Puebla, Pue., 72570, Mexico}
\affiliation{NSCL and Dept. of Physics and Astronomy, Michigan State
University - East Lansing, Michigan 48824-1321, USA}

\begin{abstract}
We propose a method to study the transition to chaos in isolated quantum systems of interacting particles. It is based on the concept of delocalization of eigenstates in the energy shell, controlled by the Gaussian form of the strength function. We show that although the fluctuations of energy levels in integrable and non-integrable systems are principally different, global properties of the eigenstates may be quite similar, provided the interaction between particles exceeds some critical value. In this case the quench dynamics can be described analytically, demonstrating the universal statistical relaxation of the systems irrespectively of whether they are integrable or not.

\end{abstract}

\pacs{05.45.Mt,05.30.-d,05.70.Ln, 02.30.Ik}
\maketitle

{\it Introduction.} - Recent experimental progresses in the studies of various quantum systems of interacting particles (see, e.g., \cite{experim}) have triggered the interest in basic problems of many-body physics. One of the issues that has been widely discussed in the literature is the onset of thermalization in an isolated quantum system caused by interparticle interactions \cite{ETH,zele,FIC96,FI97,BGIC98,I01,rigol,lea,recent}.

A prerequisite for thermalization is the statistical relaxation of the system to some kind of equilibrium
and its viability has been associated with the onset of {\it quantum chaos}. The latter term was originally created to address specific properties of dynamical quantum systems whose classical counterparts are chaotic. Later, it was found that similar properties of spectra, eigenstates and dynamics could emerge in quantum systems without a classical limit, as well as in quantum systems with disordered potentials. Nowadays, the term quantum chaos is used in a broader context when referring to those properties, irrespectively of the existence of a classical limit.

According to studies of isolated quantum many-body systems, their eigenfunctions (EFs) in the mean-field basis spread as the interaction between particles (or quasi-particles) increases, and they may eventually become {\it chaotic eigenstates} \cite{C85}. A crucial point is that due to the finite range of the interactions, only part of the {\it unperturbed} basis states $|n \rangle $ is directly coupled and therefore accessible to the eigenstates $|\alpha \rangle =\sum_{n} C^{\alpha}_{n} |n\rangle $. In the energy representation, this fraction constitutes the {\it energy shell} of the system, being either partly or fully filled by the actual eigenstates~\cite{CCGI93}. In the first case, the eigenstates are {\it localized}, having a small number of non-zero elements  $C^{\alpha}_{n}$. In the second case, when the energy shell is fully filled, the eigenstates can be either {\it sparse} or {\it ergodic} \cite{QCC}, both showing a very large number of principal components ($N_{pc} >> 1$) strongly fluctuating with $n$. In ergodic eigenstates the coefficients $C^{\alpha}_{n}$ become random variables following a Gaussian distribution around the ``envelope" defined by the energy shell. The latter scenario is that of chaotic eigenstates and occurs when the interaction exceeds a critical value~\cite{CCGI93,zele,FIC96,FI97}.

As argued in Ref.~\cite{CCGI93}, the energy shell is associated with the limiting form of the {\it strength function} (SF) for a given interaction strength. This function is defined via the projection of unperturbed states onto the basis of perturbed (exact) eigenstates. Written in the energy representation, SF is widely used in nuclear physics and is analogous to {\it local density of states} in solid state physics. It has also been shown (see, e.g., \cite{zele,FI00,FI97,K01}) that the shape of SF changes from Breit-Wigner (Lorentzian) to Gaussian as the interparticle interaction increases.

It should be stressed that if a quantum system has a classical limit, the shapes of both EFs and SFs in the energy representation have classical analogs. The quantum-classical correspondence of EFs and SFs has been studied in great detail for various few and many-body systems (see Ref.~\cite{QCC}). Typically, delocalization of EFs in the energy shell is directly related to the {\it chaotization} of the system in the classical limit, thus providing a tool to reveal the transition to quantum chaos.

The emergence of chaotic eigenstates has been related to the onset of thermalization in many-body systems, even if the latter are isolated from a heat bath \cite{ETH,zele,FIC96,FI97,BGIC98,I01,rigol,lea}. An essential point of Refs.~\cite{FIC96,FI97,BGIC98} is that when the eigenstates become chaotic, the distribution of occupation numbers achieves a Fermi-Dirac or a Bose-Einstein form, thus allowing for the introduction of temperature. Using a two-body random matrix model, a relation between temperature and interaction strength was analytically derived \cite{FI97}. Therefore, the interparticle interaction plays the role of a heat bath when the eigenstates are chaotic. Since the components of chaotic eigenstates can be treated as random variables, the eigenstates close in energy are statistically similar. This fact has been employed to justify the agreement between the expectation values of few-body observables and the predictions from a microcanonical ensemble \cite{ETH,rigol,lea}.

In this Letter we present a unified approach to identify the emergence of chaotic eigenstates. Two models of interacting spins-1/2 are considered: one model is completely integrable for any value of the interaction and the other is not. The different procedures used to determine the critical parameters above which the eigenstates become chaotic lead all to the same values. Our results reveal a relation between energy level statistics, the structure of EFs and SFs, and the quench dynamics. We show that statistical relaxation occurs for both models and can be described analytically with the same expression.

%%%%%%%%%%%%%%%%%%%%%%%%%%%%%%%%%%%%%%%%%%%%%%%%%%%%%%%%%%%%%%%%%%%%%%%%%%%%

{\it The models.} We consider two models of interacting spins-1/2. One model has only nearest-neighbor couplings, which results in the complete integrability of the system. The other case has additional next-nearest-neighbor couplings, and becomes chaotic when the two coupling strengths are comparable. Assuming open boundary conditions, the Hamiltonians read as
\begin{eqnarray}
&& H_1 = H_0 + \mu V_1 , \; \;\;\; H_2 = H_1 + \lambda V_2 ,
\label{model1} \\
&& H_0 = J \sum_{i=1}^{L-1}  \left(S_i^x S_{i+1}^x + S_i^y S_{i+1}^y \right) , \;\;\;\; V_1 = J \sum_{i=1}^{L-1}   S_i^z S_{i+1}^z,
\nonumber \\
&&
V_2 = \sum_{i=1}^{L-2} J\left[ \left( S_i^x S_{i+2}^x + S_i^y S_{i+2}^y
\right) + \mu S_i^z S_{i+2}^z  \right], \nonumber
\end{eqnarray}
where $\mu $ and $\lambda$ control the perturbation in Model 1 and  Model 2, respectively. Here, $L$ is the number of sites, $S^{x,y,z}_i = \sigma^{x,y,z}_i/2$ are the spin operators at site $i$, with $\sigma^{x,y,z}_i$ as the Pauli matrices and $\hbar=1$. The coupling strength $J$ defines the energy scale and is set to 1. We refer to a spin ``up" in the $z$ direction as an excitation, assuming the presence of a magnetic field.

In Model 1, the term $H_0$ determines the (mean field) basis in which the total Hamiltonian $H_1$ is presented.
This term moves the excitations through the chain and can be mapped onto a system of noninteracting spinless fermions~\cite{JW28} or of hard-core bosons~\cite{HP40}, being therefore integrable. The system remains integrable even when the Ising interaction $V_1$ is added, no matter how large the anisotropy parameter $\mu $ is. The total Hamiltonian $H_1$ is known as the XXZ Hamiltonian and can be solved exactly via the Bethe Ansatz~\cite{B31}.

The integrable XXZ Hamiltonian determines the mean field basis for Model 2. Thus, $V_2$ is treated as the ``residual interaction" responsible for the onset of chaos. The parameter $\lambda$ refers to the ratio between the next-nearest-neighbor and the nearest-neighbor exchange.

Depending on the parameters of the Hamiltonians (\ref{model1}), different symmetries are identified~\cite{BSSV08}. For the sake of generality, we avoid them by restricting our analysis to a subspace with $L/3$ up-spins and $\mu \neq 1$. The only remaining symmetry is parity. We take it into account by studying only even states, which leads to subspaces of dimension $N \sim (1/2) L!/[(L/3)!(L-L/3)!]$. All data are given for $L=15$ and $\mu=0.5$ for Model 2.

\begin{figure}[!ht]
\vspace{0cm}
\includegraphics[width=7cm]{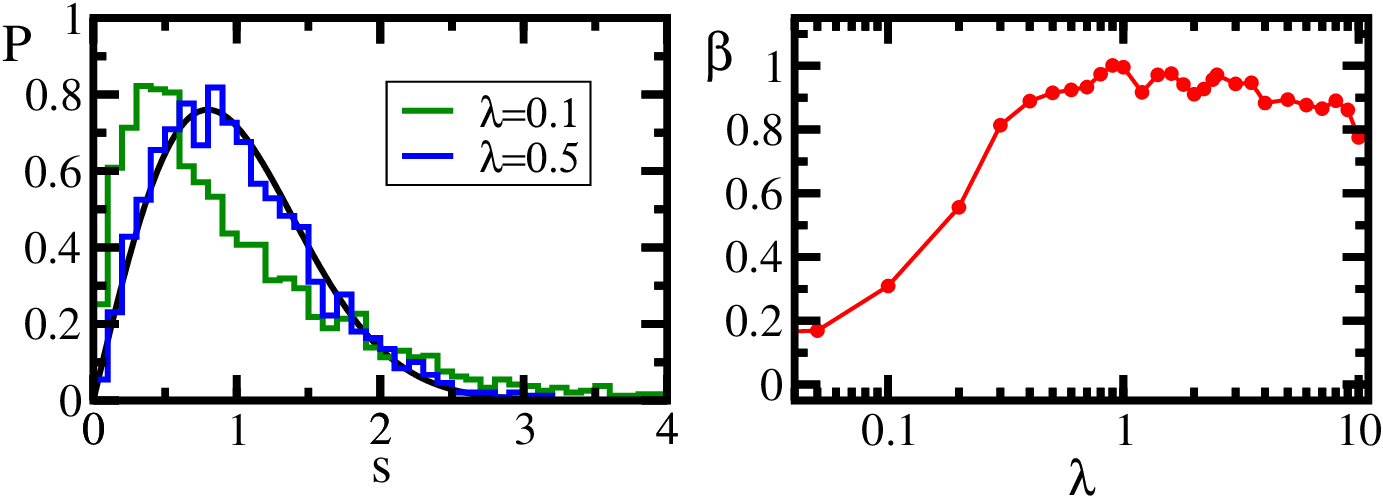}
\caption{Left: $P(s)$ for Model 2 with  $\lambda=0.1,\,0.5$  compared with the Wigner-Dyson distribution (smooth curve). Right: Brody parameter $\beta$ as a function of $\lambda$.}
\label{prima}
\end{figure}

{\it Spectrum statistics.} - According to the common lore, we analyze first the level spacing distribution $P(s)$ for both models, numerically obtained for different values of the control parameters $\mu$ and $\lambda$. For Model 1, $P(s)$ is close to the Poisson distribution for any value of $\mu$. For Model 2, the transition of $P(s)$ from Poisson to Wigner-Dyson as $\lambda$ increases is shown in Fig.~\ref{prima}. The standard approach of fitting $P(s)$ with the Brody distribution \cite{B73} allows us to extract the repulsion parameter $\beta$ characterizing the transition between the two distributions. From Fig.~\ref{prima}, the transition for Model 2 with $L=15$ occurs approximately at $\lambda \approx 0.5$.

%%%%%%%%%%%%%%%%%%%%%%%%%%%%%%%%%%%%%%%%%%%%%%%%%%%%%%%%%%%%%%%%%%%%%%%%%%%%%%%%%%%%%%%%
%%%%%%%%%%%%%%%%%%%%%%%%%%%%%%%%%%%%%%%%%%%%%%%%%%%%%%%%%%%%%%%%%%%%%%%%%%%%%%%%%%%%%%%%

\begin{figure}[!ht]
\includegraphics[width=7cm]{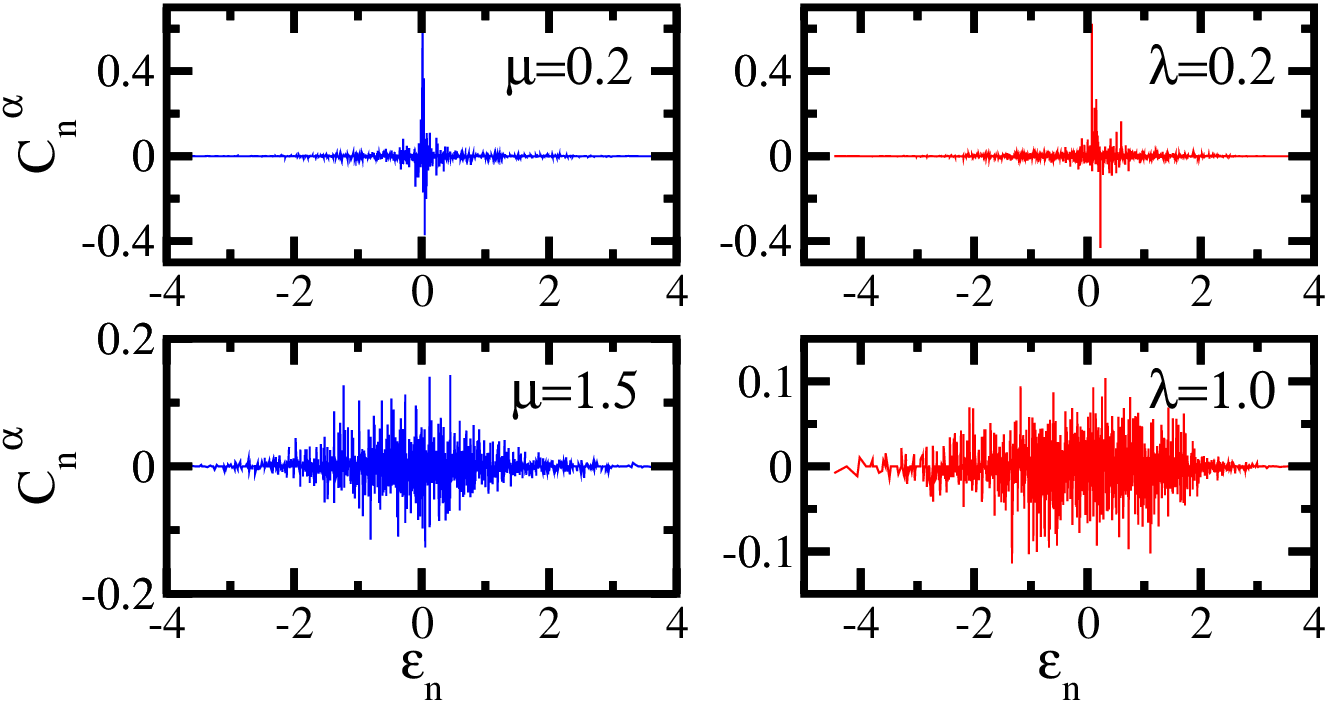}
\caption{Typical localized (top) and extended (bottom) eigenstates for Model 1 (left) and Model 2 (right).}
\label{EFs}
\end{figure}

{\it Emergence of chaotic eigenstates.} - Much more information is contained in the structure of the eigenstates. Our data show that as the strength of the perturbations $V_1$ and $V_2$ increases, the eigenstates of {\it both} integrable and non-integrable models undergo a transition from localized to chaotic-like ones. Typical examples of the amplitudes $ C_n^\alpha $ of such eigenstates with energy $E_{\alpha}$ from the center of the energy band are shown in Fig.~\ref{EFs}. Here, the eigenstates are given as a function of the unperturbed energy $\varepsilon_n$ rather than in the basis representation, following the one-to-one correspondence between each unperturbed state $|n \rangle $ and its energy $\varepsilon_n$. The figure shows how the eigenstates spread as the interaction increases.

In order to find the critical parameters $\mu_{cr}$ and $\lambda_{cr}$ above which the perturbation is strong and the eigenstates are extended (in energy shell), different approaches may be employed. We start by analyzing the matrix elements of $H_1$ and $H_2$. It is important to take into account that in each line $n$ of the Hamiltonians, the perturbation couples directly only $M_n$ unperturbed states \cite{AGKL97,FI97}. We have numerically found that at the center of the energy band, $M_n \approx N/2$ and $M_n \approx N$ for Models 1 and 2, respectively. Thus, the Hamiltonian matrix of the integrable model has more vanishing elements $H_{nm}$, being more sparse than the matrix of Model 2.

To determine the critical perturbation, we compare the average value of the coupling strength, $v_n = \sum_{m\neq n} |H_{nm}|/M_{n}$, of each line $n$ with the mean level spacing  $d_n$ between directly coupled states. The mean level spacing can be estimated as $d_n=[\varepsilon_n^{max}-\varepsilon_n^{min}]/M_{n}$, where $\varepsilon_n^{max}$ ($\varepsilon_n^{min}$) is the unperturbed energy corresponding to the largest (smallest) $m$ where $H_{nm}\neq0$. Our results show that for $\mu > \mu_{cr} \approx 0.5$ and $\lambda > \lambda_{cr} \approx 0.5 $ the ratio $v_n / d_n$ becomes larger than 1 and the perturbation is considered to be strong. Notice that for Model 2, the obtained value of $\lambda_{cr}$ corresponds to the onset of the Wigner-Dyson statistics, as  independently found from the level spacing distribution. This is remarkable if we take into account that no diagonalization was necessary to derive the above estimates.

%%%%%%%%%%%%%%%%%%%%%%%%%%%%%%%%%%%%%%%%%%%%%%%%%%%%%%%%%%%%%%%%%

{\it Strength function: From Breit-Wigner to Gaussian.} - Another way to obtain the critical values $\mu_{cr}$ and $\lambda_{cr}$ relies on the shape of SF. The latter corresponds to the dependence of $w_n^{\alpha}=|C^{\alpha}_n|^2$ on the exact energies $E_{\alpha}$ for each fixed unperturbed energy $\varepsilon_n$. It contains information about the energies $E_{\alpha}$ that become accessible to an initial state $|n \rangle $ when the perturbation is turned on. Clearly, SF is related also to the structure of EFs, since the latter is derived from the same $w_n^{\alpha}$, but now as a function of the unperturbed energies $\varepsilon_n$.

In quantum many-body systems, the form of SF typically changes from Breit-Wigner to Gaussian as the inter-particle interactions increase \cite{zele,FI00,FI97}. This transition occurs when the half-width of the Breit-Wigner distribution becomes comparable to the width of the energy shell. In this case, as we show next, there emerge chaotic eigenstates filling the whole available energy shell.

The energy shell corresponds to the distribution of states obtained from a matrix filled only with the off-diagonal elements of the perturbation. It is associated with the maximal SF, that is the SF that arises when the diagonal part of the Hamiltonian can be neglected. We verified that the energy shell coincides with a Gaussian of variance $\sigma^2$ given by the second moment of the off-diagonal elements of the matrix Hamiltonian, $\sigma^2=\sum_{m\neq n} |H_{nm}|^2$ \cite{FI97}. Note that no diagonalization is required to derive this expression.

%%%%%%%%%%%%%%%%%%%%%%%%%%%%%%%%%%%%%%%%%%%%%%%%%%%%%%%%%%%%%%%%%%%%
\begin{figure}[htb]
\includegraphics[width=7.0cm]{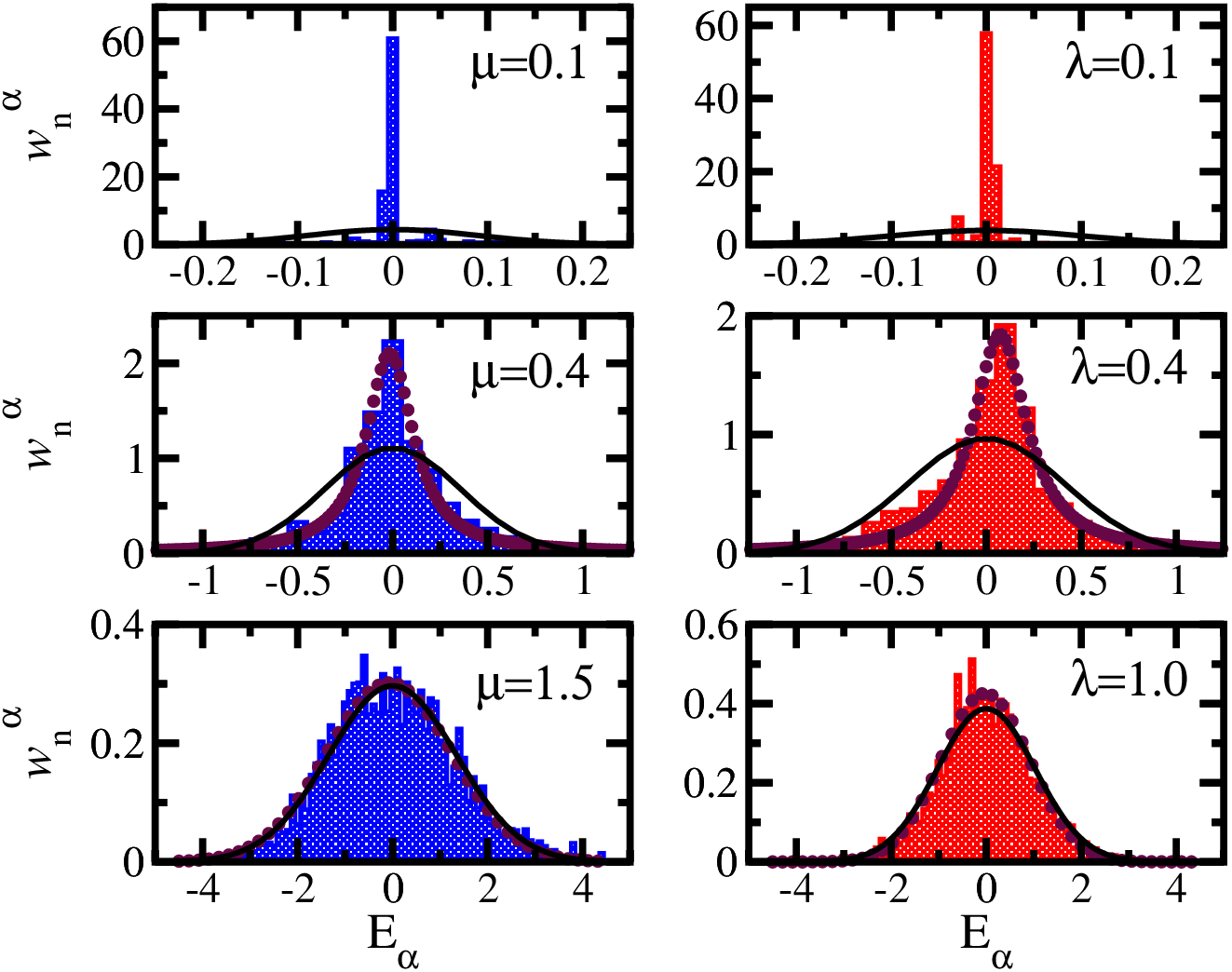}
\caption{(Color online) Strength functions for Model 1 (left) and Model 2 (right) obtained by averaging over 5 states in the middle of the spectrum. Circles give a Breit-Wigner fit ($\mu = \lambda = 0.4$) and a Gaussian fit ($\mu = 1.5$ and $\lambda = 1.0$). Solid curves correspond to the Gaussian form of the energy shells.
}
\label{le}
\end{figure}
%%%%%%%%%%%%%%%%%%%%%%%%%%%%%%%%%%%%%%%%%%%%%%%%%%%%%%%%%%%%%%%%%%%%%%%%%%%%%

Our numerical data confirm that the transition from Breit-Wigner to Gaussian occurs for the same critical values obtained above, $\mu_{cr} \, , \lambda_{cr} \approx 0.5$, as indicated in Fig.~\ref{le}.
In the figure, the envelopes of the SFs were obtained by smoothing the dependence of $w_n^{\alpha}$ on $E_{\alpha}$  for fixed unperturbed energies $\varepsilon_{n}$ with $n \approx N/2$. The fit to either a Breit-Wigner or a Gaussian form was done with high accuracy, allowing us to discriminate between the two functions. It is noteworthy the excellent agreement between the Gaussian fit and the Gaussian obtained simply from the off-diagonal elements of the Hamiltonians.

%%%%%%%%%%%%%%%%%%%%%%%%%%%%%%%%%%%%%%%%%%%%%%%%%%%%%%%%%%%%%%%%%%%%%%%%%%%%%%%%%

\begin{figure}[htb]
\includegraphics[width=7.0cm]{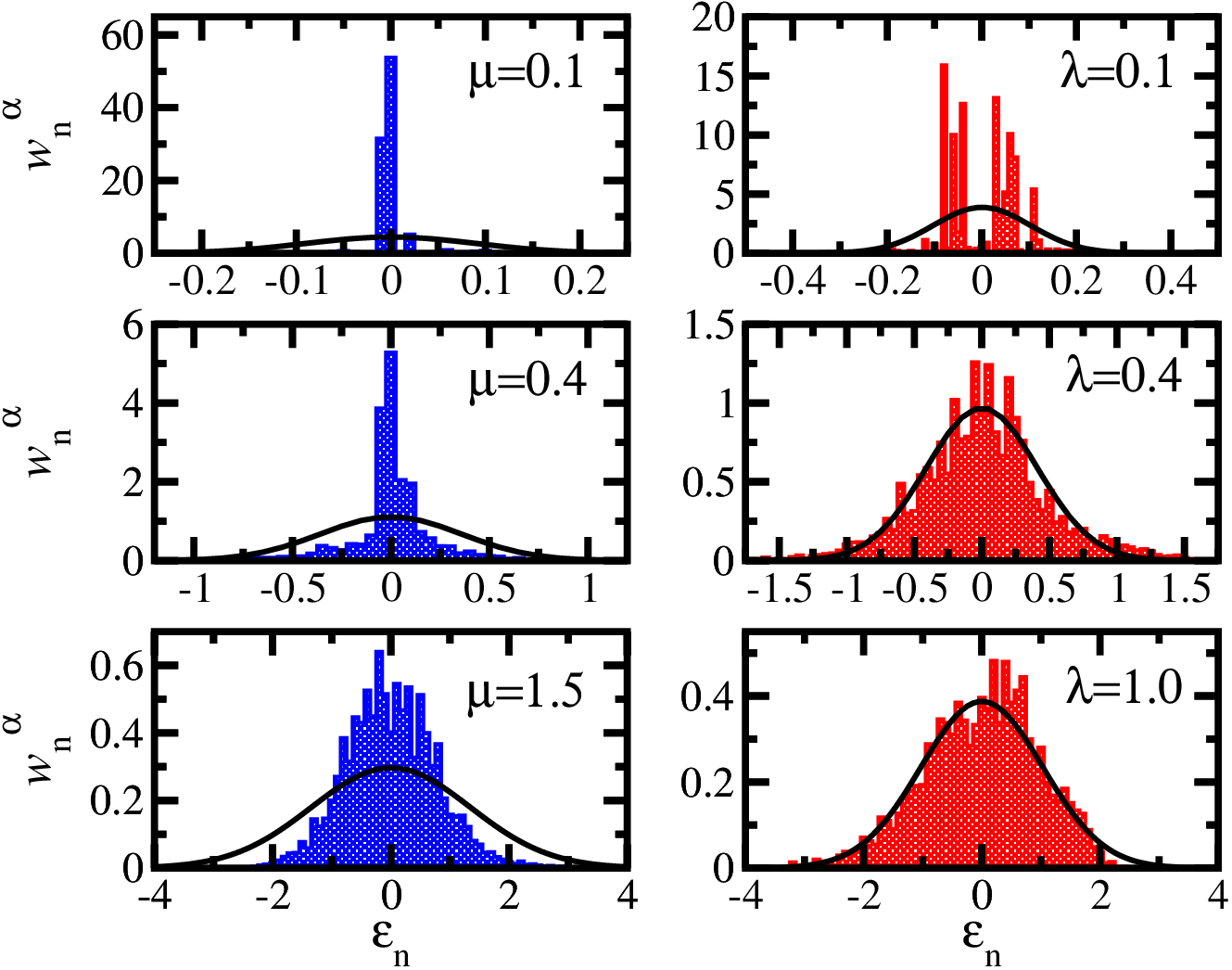}
\caption{Structure of eigenstates in the energy shells for Model 1 (left) and Model 2 (right) obtained by averaging over 5 states in the middle of the energy band. Solid curves correspond to the Gaussian form of the energy shells.
}
\label{new}
\end{figure}

{\it Structure of eigenstates in energy shell.} - The eigenstates may be localized, sparse or ergodically extended in the energy shell. The data in Fig.~\ref{new} demonstrate that for a sufficiently strong perturbation the eigenstates undergo a transition from strongly localized to extended states, somehow filling the energy shell. The transition to chaotic eigenstates occurs again at the same critical parameters $\mu_{cr} \, , \lambda_{cr} \approx 0.5$. These results confirm the predictions made on the basis of both, the estimate of $v_n/d_n$ and the Gaussian form of the strength functions.

One notices that above the critical value, the eigenstates of Model 2 appear to fill the energy shell better than those of Model 1 (cf. bottom panels of Fig.~\ref{new}). Indeed, for Model 1, the eigenstates are not completely extended in the energy shell even for strong interaction, $ \mu =1.5$, although they do fill a large part of it. At the same token, a close inspection of the level of delocalization of individual EFs and SFs has revealed differences between the two models. Overall, delocalization measures, such as the inverse participation ratio or Shannon entropy, show larger fluctuations for Model 1 than for Model 2 \cite{future}. This agrees with recent results obtained for bosonic and fermionic systems \cite{lea}.

%%%%%%%%%%%%%%%%%%%%%%%%%%%%%%%%%%%%%%%%%%%%%%%%%%%%%%%%%%%%%%%%%%%%%%%%%%%%%%%%%

{\it Statistical relaxation.} - The knowledge of the shape of SFs allows one to describe the quench dynamics in a system of interacting particles, provided the eigenstates are chaotic and delocalized in the energy shell \cite{FI01}. By quench dynamics we mean the time evolution of initial states corresponding to unperturbed basis vectors which takes place once the interaction is turned on. To see how the relaxation occurs, we study the time dependence of the Shannon entropy $S$ in the unperturbed basis. For an initial state $|k\rangle$, it is defined as
\begin{equation}
S_k(t)=-\sum_{n=1}^N  \Omega_n^{(k)} (t) \ln \Omega_n^{(k)} (t)
\label{entropy}
\end{equation}
with
$\Omega_n^{(k)} (t) = |\sum_{\alpha} C^{\alpha*}_k C^{\alpha}_n e^{-iE_{\alpha}t}|^2$.

\begin{figure}[htb]
\includegraphics[width=0.4\textwidth]{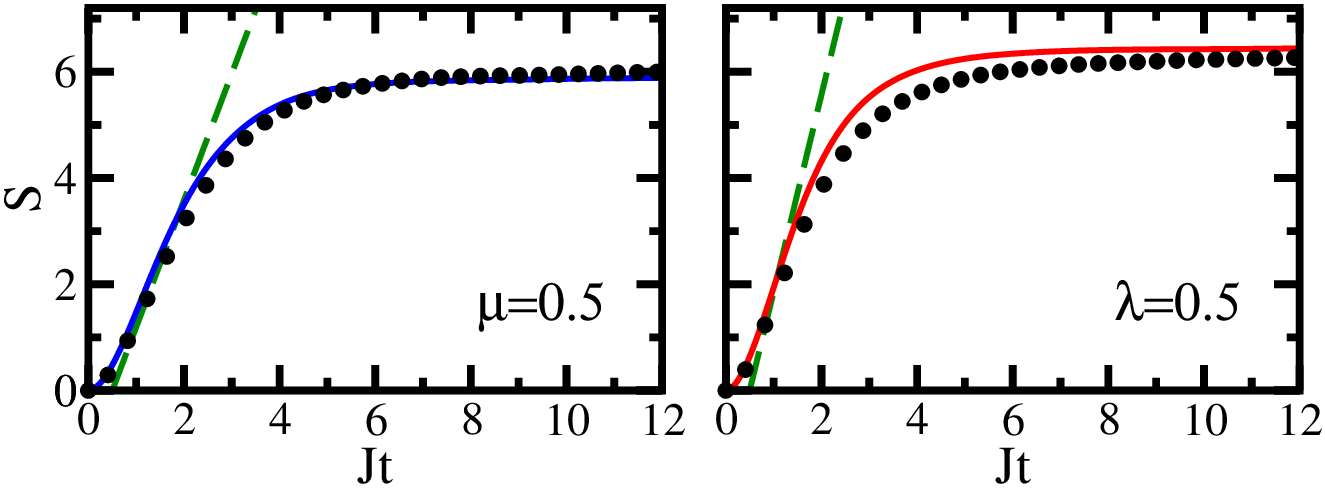}
\caption{(Color online) Shannon entropy {\it vs} rescaled  time for Model 1 (left) and Model 2 (right) for strong perturbation. Circles stand for numerical data, solid curves correspond to Eq.~(\ref{analyt}), and dashed lines show the linear dependence (\ref{linear}).}
\label{fig:analytics}
\end{figure}

An analytical expression for $ S{(t)}$ has been derived with the use of a two-body random matrix model \cite{FI01},
\begin{equation}
S_k(t) = -W_k(t) \ln W_k(t) -
[1-W_k(t)] \ln \left( \frac{1-W_k(t)}{N_{pc}} \right).
\label{analyt}
\end{equation}
Here $W_{k}(t)$ is the probability for the system to stay in the initial state $|k \rangle $ and $N_{pc}$ is the average number of directly coupled states. We obtain $N_{pc}$ numerically according to $N_{pc}=\langle e^S \rangle$, where the average $\langle. \rangle$ is performed over a long time after the entropy saturates. As for $W_{k}(t)$, it has been shown \cite{FI00} that for a Gaussian SF, it decays as $W_{k}(t)= \exp (-\sigma^2 t^2)$, with $\sigma^2$ as previously defined.

Figure \ref{fig:analytics} shows numerical data for the relaxation process of both models. To reduce fluctuations, we average over 5 initial basis states excited in a narrow energy range in the middle of the spectrum. Initially, the entropy grows quadratically, as given by perturbation theory. Afterwards, a clear linear growth is observed before $S(t)$ reaches relaxation. With high accuracy the linear behavior of $S(t)$ is described by the simple relation \cite{FI01},
\begin{equation}
S_k(t) \approx \sigma_k t \ln M_k.
\label{linear}
\end{equation}
Note that Eq.~(\ref{linear}) depends only on the elements of the Hamiltonian: $\sigma_k^2=\sum_{m\neq k} |H_{km}|^2$ and $M_k$ is the connectivity, that is the number of directly coupled unperturbed states in the
$k$-th line of the Hamiltonian matrix. As seen in Fig.~\ref{fig:analytics}, the analytical expressions (\ref{analyt}) and (\ref{linear}) give a correct description of the increase of the entropy for {\em both} models in the regime corresponding to the onset of chaotic-like eigenstates delocalized in the energy shell. The same relation (\ref{linear}) was found to emerge also for an integrable model of interacting bosons \cite{BBIS03}.

{\it Conclusion.} - We have studied spectrum statistics,   structures of eigenstates and strength functions for two models of interacting spins, connecting the results with the onset of chaotic-like eigenstates and the statistical relaxation of the quench dynamics. The key point of our approach is the existence of an energy shell of finite range, inside which the eigenfunctions can be either localized or extended. We have shown that the critical parameters above which the eigenstates become chaotic can be equally found by simply studying the elements of the Hamiltonian or by analyzing the shape of the strength function. The latter provides us with the form of the energy shell, thus allowing one to clearly define the notion of delocalized eigenstates in the energy shell.

By studying the time dependence of the Shannon entropy, we have shown that numerical data are in full agreement with the analytical predictions of the quench dynamics, provided the eigenstates are chaotic-like.
Our approach is very general and expected to apply to different systems of interacting
particles, such as those currently under theoretical and experimental investigation.

{\it Acknowledgments.} F.M.I. acknowledges support from CONACyT grant N-161665 and thanks Yeshiva University for the hospitality during his stay in the fall 2011.

\end{document}